# Stress and charge transfer in uniaxially strained CVD graphene

**Milan Bousa** [*,1,2], **George Anagnostopoulos**[3], **Elena del Corro**[1], **Karolina Drogowska**[1], **Jan Pekarek**[4], **Ladislav Kavan**[1,2], **Martin Kalbac**[1], **John Parthenios**[3], **Konstantinos Papagelis**[3], **Costas Galiotis**[3,5], and **Otakar Frank**[1]

[1] J.Heyrovsky Institute of Physical Chemistry of the AS CR v.v.i., Dolejskova 2155/3, CZ 182 23 Prague 8, Czech Republic
[2] Department of Inorganic Chemistry, Faculty of Science, Charles University in Prague, Albertov 6, CZ 128 43 Prague 2, Czech Republic
[3] Institute of Chemical Engineering Sciences, Foundation for Research and Technology—Hellas (FORTH/ICE-HT), P.O. Box 1414, Patras 265 04, Greece
[4] Centre of Sensors, Information and Communication Systems, Faculty of Electrical Engineering and Communication, Brno University of Technology, Technicka 3058/10, CZ 616 00 Brno, Czech Republic
[5] Department of Chemical Engineering, University of Patras, Patras 26504, Greece

Mechanical properties of graphene prepared by chemical vapor deposition (CVD) are not easily comparable to the properties of nearly perfect graphene prepared by mechanical cleavage. In this work, we attempt to investigate the mechanical performance of CVD graphene (simply supported or embedded in polymer matrix), transferred by two different techniques, under uniaxial loading with simultaneous in-situ monitoring by Raman microspectroscopy. The level of charge transfer doping and strain is assessed using the vector analysis modified for uniaxial strain. The strain distribution across the samples varies significantly, owing to the growth and transfer process, which induces wrinkles and faults in the CVD graphene.

In simply supported specimens, the stress transfer efficiency is generally very low and the changes in Raman spectra are dominated by variations in the charge transfer originating from the realignment of the domains on the substrate upon the application of strain. In contrast, samples covered with an additional polymer layer exhibit an improved stress transfer efficiency, and the alterations of charge doping levels are negligible. In fully embedded specimens, the variations in stress transfer efficiencies are caused by the size of the effective graphene domains defined by cracks, folds and/or wrinkles.

**1 Introduction** The presence of strain [1] and substrate-induced charge transfer doping [2] in graphene requires reliable methods for their monitoring and correct evaluation. Both strain and doping are known to alter the electronic structure of graphene, intentionally [3] as well as accidentally [4]. Hence, the ability to correctly assess and discriminate strain and doping in graphene and graphene-based devices is of utmost importance.

Raman spectroscopy has been established as a work horse for a swift and non-destructive analysis of graphene-related materials, providing not only basic characterization in terms of the layer number or disorder in the samples, but also a more detailed information about the levels of doping or strain [5, 6]. Nevertheless, when these two effects intermix, such an analysis can become very difficult. A method using the correlation of the G and 2D peak frequencies (Pos(G) and Pos(2D), respectively) has been introduced [7] and successfully tested recently [8-11] to separate biaxial strain from charge doping in various graphene samples. This approach, based on a vector analysis of the data points in the Pos(G)-Pos(2D) space, is relatively simple to conduct in certain well-defined cases, but the interpretation of the results has to be performed carefully when the experimental conditions deviate from the above mentioned biaxial strain-uniform doping situation. In that case, to monitor the changes in the system, other parameters of the G and 2D peaks have to be analyzed, such as their widths (further defined as full-width at half-maxima, FWHM), or their intensity ratio [8]. Alternatively, another Raman feature, such as the 2D' peak can be included to provide more accuracy in the analysis [12].

The G peak corresponds to a first order Raman scattering process with a phonon of almost zero momentum. It is associated with the in-plane, doubly degenerate phonon from the transverse optical (TO) and longitudinal optical (LO) branches with $E_{2g}$ symmetry at the Brillouin zone (BZ) center ($\Gamma$ point) [13]. Both frequency and width of the G mode are strongly influenced by the presence of doping [14, 15] and the application of stress [1, 16]. The increase of Pos(G) in both electron- and hole-doped graphene is caused by a non-adiabatic removal of the Kohn anomaly at $\Gamma$ point [17], and the simultaneous decrease of FWHM(G) is caused by the Pauli blocking of phonon decay into electron-hole pairs. In general, strain causes G peak downshift under tension and upshift under compression with a rate of ~62 cm$^{-1}$/% for biaxial tension [18]. Under uniaxial strain the G peaks splits into two components, G$^-$ and G$^+$, with the shift rates of ~ -31 and -10 cm$^{-1}$/%, respectively, for graphene on polymer substrates [19, 20], or of ~ -37 and -19 cm$^{-1}$/%, resp., for suspended graphene [21].

The 2D mode originates from a second-order triple resonant process between non-equivalent K points in graphene BZ, involving two zone-boundary, TO-derived phonons with opposite momenta [22, 23]. As shown recently, the 2D peak in suspended graphene is not fully symmetrical [24], which is even more pronounced in strained samples [25-28]. In a simplified one-dimensional picture of the 2D mode origin there are two dominant directions of the contributing phonon wavevectors – along K-$\Gamma$ (so called inner) or K-M (outer) symmetry lines. Recent studies show a greater contribution of the inner processes [22, 25, 26, 29]. Nevertheless, a full two-dimensional description of the electronic peaks, phonon dispersion and matrix elements [22] has shown that the notation of inner or outer phonons is of a weaker relevance [28]. The 2D mode is dispersive, and its frequency changes with the excitation energy ($E_L$) with the slope $\partial Pos(2D)/\partial E_L$ ~ 100 cm$^{-1}$/eV [6]. The 2D peak is also sensitive to doping and mechanical stress but these effects manifest themselves differently from those of the G peak [1]. Strain causes the 2D peak shift in the same directions as the G peak, with the shift rates for biaxial strain larger by a factor of ~ 2.2-2.5 [18, 30, 31]. Broadening and splitting of the 2D peak under uniaxial strain is very different to that of the G peak [25], but no such effects are observed for biaxial deformation [31]. Hole doping causes increase of Pos(2D) with a $\partial Pos(2D)/\partial Pos(G)$ ~ 0.5-0.7 [7, 14, 32], whereas electron doping causes only a negligible Pos(2D) change for $n \leq 2 \times 10^{13}$ cm$^{-2}$ followed by a non-linear Pos(2D) decrease for higher n-doping levels [14, 32]. FWHM(2D) increases upon both p- and n-doping mainly due to electron-electron interactions, and also electron-phonon coupling strength [22, 33].

The effect of the grain size on the stress transfer was reported for mechanically exfoliated graphene, showing that a critical dimension of approximately 4 μm (parallel with the strain axis), is needed to achieve a full transfer of the applied strain from the substrate to the interior of the flake [34-36]. The strain-induced behaviour of CVD graphene is much less explored [37, 38], and the influence of absolute grain sizes is taken over by the wrinkles, which effectively rule the stress transfer, or rather the lack thereof, in this case [37]. The shift rate of 2D peak with strain was less than 25% of the shift rate corresponding to a full stress transfer (~60 cm$^{-1}$/%), in spite of the flake being covered by an additional polymer layer [37]. The analysis of the G peak was not possible due to intense overlapping peaks originating from the poly(ethylene terephthalate) (PET) substrate.

In the present work, we have focused on the separation of the effects of strain and doping in the Raman spectra of uniaxially strained CVD graphene prepared by two different transfer methods: (i) with separated grains introduced intentionally using an elastomer sacrificial layer (polyisobutylene, PIB) and (ii) the standard poly(methylmethacrylate) (PMMA) technique. Apart from confirming the influence of grain sizes, both absolute and "wrinkle-limited", we have shown the applicability of the strain-doping separation method also for uniaxially strained graphene samples.

**2 Methods** Graphene was grown on copper foils by low-pressure CVD as detailed in Ref. [39]. Alternatively, single layer graphene produced by CVD method and grown on copper foil supplied by AIXTRON Ltd was used in the experiments.

The samples were transferred either by the dry transfer method using PIB as support polymer [40] or the standard PMMA method [41]. Scanning Electron Microscopy (SEM) images of the samples transferred by the two methods are shown in Figure 1. The images were acquired using a HR-SEM Hitachi S4800 and a LEO SUPRA 35 VP scanning electron microscopes. In the case of PIB-assisted transfer, the SEM images were obtained in samples transfer to Si substrate due to a lower conductivity of the monolayer caused by the fragmentation during transfer.

As the target substrate for tensile experiments, clean and flexible polymethylmethacrylate (PMMA) beams of 3 mm thickness were used. They were previously spin coated with different polymers, SU8 photoresist (SU8 2000.5, MicroChem) or PMMA (3% in anisole, MicroChem). After the graphene transfer, some of the samples were covered by parylene C (120 nm thickness) or PMMA, in order to improve the strain transfer efficiency from the flexible substrate to the graphene flake. Further in text, the uncovered samples are denoted as simply supported and the covered ones as fully supported. Tensile strain in the graphene was imposed through a cantilever beam [19] or a four-point bending device [36]. In both cases, the strain response was monitored in-situ using a micro-Raman setup with a 100x objective (N.A.=0.9) (InVia Reflex, Renishaw, or LabRAM HR, Horiba), and 514.5 nm (2.41 eV) or 632.8 nm (1.96 eV) excitation laser. The laser power was kept below 1.5 mW to avoid laser-induced heating. Typical spectra of single layer CVD graphene, as transferred onto SU8 and strained, are plotted in Figure 2.

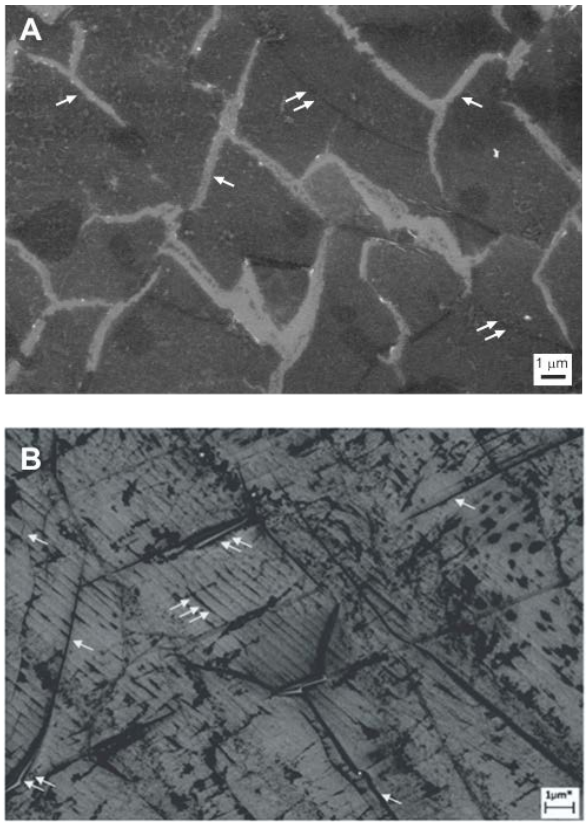

**Figure 1** SEM images of the CVD graphene samples transferred by PIB (A; on Si) and PMMA (B; on SU8). The scale bar is 1 μm.

**3 Results** Figures 2 and 3 show the evolution of the monolayer graphene flake covered by parylene under uniaxial strain. In the Raman spectra (Fig. 2), downshifts of both the G and 2D peaks of graphene are visible in the spectrum strained to $\varepsilon_m = 0.5$ %. The Raman peaks of the polymers do not shift with strain. The measurement was done using incident laser polarization parallel to the strain axis; no selection of the scattered light polarization (via an "analyzer") was employed. The graphene Raman peaks thus correspond to a sum of signals from all scattered polarization directions and the shift rates can be taken from single Lorentzian shapes fitted to the Raman peaks. The evolution of the Raman shifts of the G and 2D peaks with strain is plotted in Fig. 3c and b, with their shift rates being 12.7 and 19.4 cm$^{-1}$/%, respectively. In the case of the 2D peak (Fig. 3b), the data points lie on a line, whereas the G peak data points are markedly more scattered and its position at 0.5% strain is even increased compared to the previous step (note, the shift rate is therefore calculated only between 0 and 0.45% of strain). The erratic behaviour of the G peak compared to the smooth evolution of the 2D peak is a first indication of another influence on the Raman spectra apart from the strain. A second indication comes from the shift rates. Compared to the theoretical peak shifts (~ 65 cm$^{-1}$/% for the 2D peak and ~ 21.5 cm$^{-1}$/% for the G peak), we can see a serious discrepancy in the ratios between the measured and theoretical values: a factor of ~ 0.33 and 0.59 for the 2D and G peaks, respectively. Note that the theoretical G band shift is the average of the shift rates for G$^+$ and G$^-$ [1].

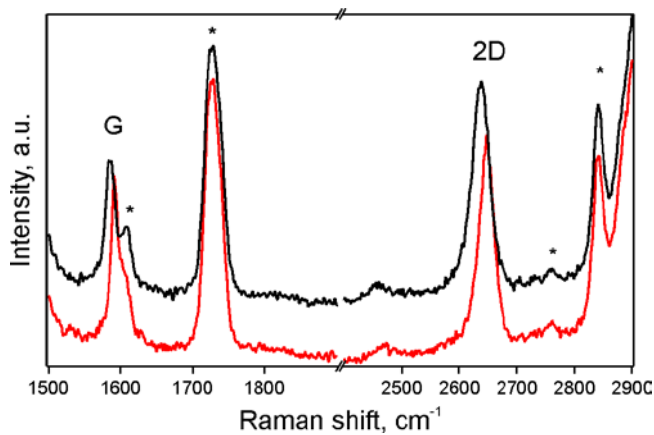

**Figure 2** Raman spectra of CVD graphene embedded between two polymer layers (SU8 at the bottom, parylene C on top) at 0 (red line) and 0.5% (black line) of nominal strain. Raman peaks of the polymer are marked by asterisks. Laser excitation energy was 1.96 eV.

Both above mentioned observations – the discrepancy between the measured and theoretical values of the G and 2D peaks and the scatter of the G peak Raman shifts - can be explained by doping. The variations in the G peak position can be rationalized by the presence of spatial fluctuations of carrier concentrations, sometimes termed as charge puddles [42], and by their fluctuation caused by the strain-induced changes in the graphene-substrate interactions. The different ratios of the G and 2D peak shifts with respect to the theoretical values can reflect a continuous change in the doping level, which influences more the G peak shift. Hence, the strain shift rates should be checked and adjusted (if needed) using the vector separation method to account for the different shifts caused by strain and doping (Fig. 3a). Apart from the point at $\varepsilon_m = 0.5\%$ strain, the evolution of the G and 2D Raman shifts at the single-spot (Fig. 3a) can be fitted by a line with the slope of ~ 1.5. This value indicates a substantial influence from changes in doping. We apply the vector analysis using the slope for strain (i.e. iso-doping) line of 3 obtained from the average peak shifts for uniaxial strain in fully supported graphene [1] and the slope for hole doping of 0.7 [7] and subsequently quantify the data using the above mentioned shift rates for strain and a simplified formula for doping estimation $\triangle\text{Pos}(G) = -0.986n^2 + 9.847n$ ($10^{13}$ cm$^{-2}$) [8], where $n$ is the carrier concentration. We obtain $\triangle n \sim -0.37\times 10^{13}$ cm$^{-2}$. The corresponding adjusted G peak strain shift rate is 4.3 cm$^{-1}$/%, amounting to ~0.2 of the theoretical shift rates.

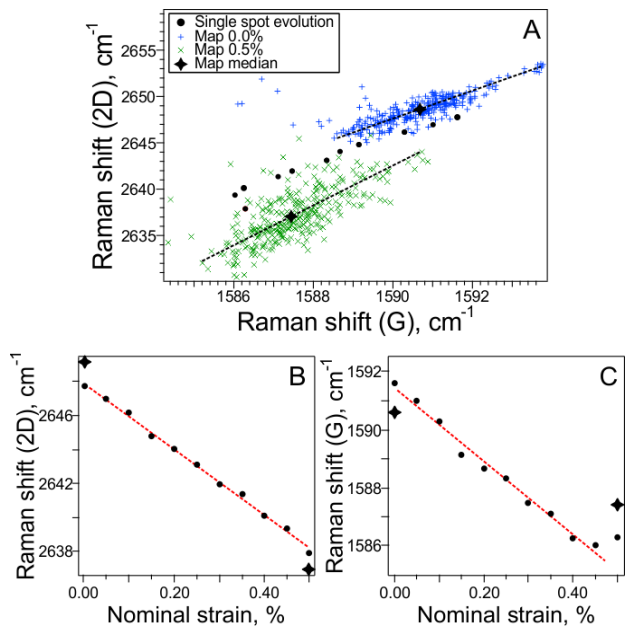

**Figure 3** Evolution of G and 2D peaks in the PIB-transferred, fully supported sample: (A) data points from large area mapping (0% - blue, 0.5% - green) with stars as their medians and full lines as least-squares line fits, black points follow the evolution on a single spot with 0.05% step, which is then detailed individually in (B), (C) for the evolution of 2D and G peaks with strain. Laser excitation energy was 1.96 eV.

As can be seen, even a small change in the doping level can lead to considerable changes in the quantified strain. The decrease in carrier concentration (dedoping) observed during the single-spot measurement is, however, counterintuitive. In general, such lower carrier concentration should be accompanied also by a loss of adhesion to the substrate, which in turn would be reflected in smaller Raman shifts of both G and 2D peak per strain step. Nevertheless, the Raman shifts – especially of the 2D peak - show a linear evolution in the whole experiment (Fig. 3b and c). It can be hypothesized that the laser spot was located in a graphene grain (charge puddle) with a higher p-doping level than the rest of sample and by flattening out the graphene's surface during the tensile test, the charge has been distributed along the neighbouring grains. Indeed, the data points acquired in maps taken at 0 and 0.5% of nominal strain (Fig. 3b) clearly evidence that the behaviour in the one spot randomly selected to follow the stretching experiment shows an anomalous evolution compared to the bulk of the data. The medians of the distributions lie on a line with a slope of ~ 3.6, indicating that even a minor overall increase in the net doping took place. The increased doping level is expected due to an increasing contact between graphene and the substrate. The vector analysis performed on the medians yields the strain G peak shift rate of ~ 8.9 cm$^{-1}$/% and $\triangle n$ of ~ 0.10×10$^{13}$ cm$^{-2}$, giving the ratio to the theoretical strain shift rate of 0.44. The ratio is higher than the one previously measured for CVD graphene on a different system with a PET as the bottom and PMMA as the top substrate [37]. If the 2D peak shift would be compared without adjusting for the doping effects, the ratio to the theoretical shift rate would give 0.38 in our case. We can also see that the distributions of the map points in Figure 3a changed their orientation: from 1.5 at 0% to 2.2 at 0.5% indicating again that upon tension the doping levels are equalized, whereas the differences between the points are dominated by strain.

The case of the fully supported CVD monolayer transferred by PMMA is depicted in Figure 4. The distribution of the mapping points changes its orientation from 1.9 to almost 2.6 at 0.8% of nominal strain (Fig. 4a), indicating the increased domination of strain. The average shift rates of the 2D and G peaks (Figs. 4b and c) are 10.4 and 3.4 cm$^{-1}$/%, resp., and their ratio ~3.1 showing a minor increase in net doping level. After an adjustment of those doping effects, the G peak shift rate increases to 3.6 cm$^{-1}$/% and the overall ratio to the theoretical shift rate is 0.17.

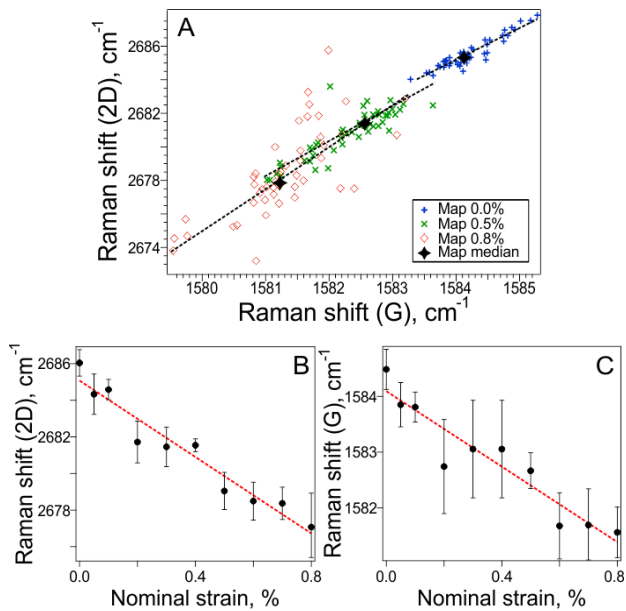

**Figure 4** Evolution of G and 2D peaks in the PMMA-transferred, fully supported sample: (A) data points from large area mapping (0% - blue, 0.5% - green, 0.8% - red) with stars as their medians and full lines as least-squares line fits, (B), (C) show the evolution of 2D and G peaks with strain. Laser excitation energy was 2.41 eV.

The reasons for difference in the shift rates between the PIB- and PMMA-transferred samples can be attributed to the effective grain sizes. The polymer-assisted transfer induces all kinds of disorder and inhomogeneities to the sample, mostly due to a larger area of the graphene film compared to the copper surface, and large surface tension when dissolving the polymer. The most common features thus formed are tears or cracks, individual folds or periodic wrinkles. In Figure 1, we can see a marked difference between the two samples. While the PIB-transferred one shows predominantly cracks and folds, the PMMA transfer retains a much better continuity of the film, but induces many wrinkles. When we measure the mean sizes of individual graphene domains separated by either of those discontinuities, we get ~2 μm for PIB transfer and less than ~0.5 μm for PMMA transfer. The larger effective domain size of the PIB-transferred samples is then reflected in the higher strain transfer efficiency.

Let us now discuss the differences between the two above mentioned cases from the viewpoint of Raman peak widths. The mean values of FWHMs of the G and 2D peaks at zero and maximum achieved strain for both the PIB- and PMMA-transferred samples are given in Table 1.

**Table 1** Raman peak widths for PIB- and PMMA-transferred, fully supported samples.

|      | Strain, %* | FWHM(G), cm$^{-1}$** | FWHM(2D), cm$^{-1}$** |
|------|------------|----------------------|------------------------|
| PIB  | 0.00       | 8.9 ± 0.8            | 31.2 ± 0.9             |
|      | 0.22       | 11.6 ± 1.2           | 33.8 ± 1.5             |
| PMMA | 0.00       | 15.7 ± 0.9           | 30.2 ± 0.7             |
|      | 0.14       | 18.6 ± 0.8           | 33.8 ± 2.4             |

* The real strain, calculated from the stress transfer efficiency. ** Average values ± standard deviation obtained from Raman maps.

In general, three major effects are responsible for changes in the Raman peak widths in monolayer graphene. Both G and 2D peaks broaden and eventually split into two (G peak) or more (2D peak) components with increasing uniaxial strain due to symmetry lowering [1]. The G peak broadening (splitting) follows a monotonic trend (in the case when scattered light with all polarization directions is registered), whereas the evolution of the 2D peak width with uniaxial strain is more complicated, non-linear and also dependent on the laser excitation energy [25]. Charge doping induces narrowing of the G peak (already at small doping levels) and broadening of the 2D peak (more pronounced for higher doping levels); neither of the effects being linear [15]. And finally, broadening of both peaks can be caused by disorder, stemming either from structural defects [43] and/or from random fluctuations of strain and doping within the laser spot [44]. As can be seen, the peak widths represent an intricate superposition of several factors, especially if uniaxial strain is present, therefore the interpretation of FWHM values has a limited usage. However, several simple assumptions can be made comparing the numbers in Table 1 and the peak shifts from Figures 3 and 4 and the discussion thereof. At zero strain level, the small FWHM(G) of the PIB-transferred sample shows a high doping level. The FWHM(G) of the PMMA-transferred sample is significantly higher, and thus smaller doping can be expected. In the same time, the FWHM(2D) of the two samples is larger than 30 cm$^{-1}$ and alike within the error margin. This indicates that apart from differences in doping, also varying amount of disorder can be traced between the two samples. Since the D band intensity is very small in both cases, the main source of disorder comes from strain/doping fluctuation in the laser spot. In the PIB sample at zero strain, the main source of 2D peak broadening comes from doping, while in the PMMA sample, the broadening originates in nanoscale strain/doping fluctuations. The latter conclusion is also in line with the much higher density of discontinuities like folds or wrinkles in the PMMA sample, as discussed above.

Upon uniaxial loading, both peaks broaden in both sample types. However, we shall consider only the evolution of the G peak width due to the complicated nature of 2D peak broadening. The theoretical rate of G peak splitting with uniaxial strain is ~ 21 cm$^{-1}$/%, a value almost identical to the average shift rate of the two components (~20.5 cm$^{-1}$/%). Hence, at smaller strains, where the two components are strongly overlapping and not discernible, one can expect the rate of G peak broadening as a function of G peak shift, $\triangle$FWHM(G)/$\triangle$Pos(G), of ~ -1 cm$^{-1}$/cm$^{-1}$. While the G peak evolution of the PMMA-transferred sample follows exactly this trend, $\triangle$FWHM(G)/$\triangle$Pos(G) in the PIB sample reaches only -0.83. The difference can be caused either by decreasing disorder in the PIB sample, or more likely – in accordance with the results obtained from the vector analysis – by a small increase of the doping level in the PIB sample, which would slightly counterbalance the strain-induced broadening. As a final note, the different doping level in the initial state of the samples seems to be reflected also in the stress-transfer efficiency – a better contact between the sample and the substrate(s) leads to a higher doping level as well as to a better stress transfer.

As another example of the usefulness of the vector analysis, a simply supported PIB-transferred CVD graphene was tested. In this case, the evolution of the G and 2D peaks in one spot show a completely different scenario, where the frequency of the 2D peak decreases, while the G peak frequency increases (Fig. 5b and c). Such a behaviour can be explained only by a commanding change in the doping level ($\triangle n$ ~ 0.27×10$^{13}$ cm$^{-2}$) with a minor contribution of a tensile strain (ratio to the theoretical shift rates of ~ 0.01). On top of that, the large spread of the values from the linear fit evidences substantial heterogeneities into the laser spot. The statistical evaluation in a larger area documents a slightly different overall evolution, which is dominated even more strongly by an increased doping level, where the medians of the two groups of data points are connected by a line with a slope of 0.7. Hence, during the experiment, most of the area of the sample experiences only a better alignment along the substrate, thereby increasing the doping level, and only in a few points the graphene is pinned to the substrate causing a minor interfacial stress transfer.

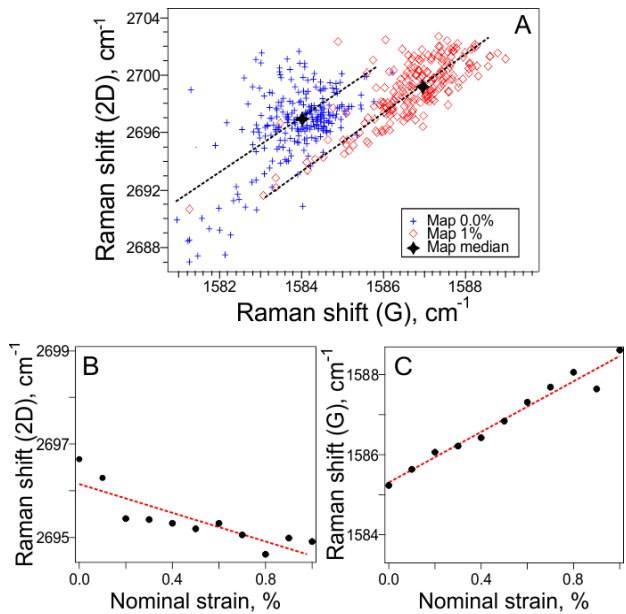

**Figure 5** Evolution of G and 2D peaks in the PIB-transferred, simply supported sample: (A) data points from large area mapping (0% - blue, 1.0% - red) with stars as their medians and full lines as least-squares line fits. (B), (C) show the evolution of 2D and G peaks with strain on a single spot. Laser excitation energy was 1.96 eV.

**Conclusions** We have shown the usability of vector analysis, performed on data points acquired by analysing the Raman G and 2D peak positions in graphene, to separate effects of strain and charge transfer doping in uniaxially strained CVD graphene samples. In all examined cases, uniaxial strain causes alignment of the graphene monolayer on the substrate resulting in an increase of doping level, however, only specimens covered by an additional polymer layer exhibit a certain amount of interfacial stress transfer. Its level is defined by the size of the effective graphene domains, which are separated by any kind of cracks, folds or wrinkles. The stress transfer efficiency is found to be ~ 44% for PIB-transferred graphene with domain sizes of ~ 2 μm and ~ 17% for PMMA-transferred sample with domain sizes of ~ 0.5 μm. A qualitative study of the G and 2D peak widths supports the conclusions from the vector analysis. It is also worth noting that single spot analysis can yield to misleading interpretation, since the samples display large strain variations and/or doping inhomogeneities. Therefore a larger area Raman mapping has to be conducted and evaluated to provide quantifiable information about the real strain and doping levels induced by the arrangement and flattening of the domains upon loading.

**Acknowledgements** This work was funded by the Graphene FET Flagship (Graphene-Based Revolutions in ICT and Beyond, Grant No. 604391). O.F., M.B and E.d.C. acknowledge the support of Czech Science Foundation project No. 14-15357S and M.K. acknowledges the support from ERC-CZ project No. LL1301. J.Pe. acknowledges the support by the project of National Sustainability Program under grant LO1401 of SIX Research Centre. Also, G.A., J.Pa., K.P. and C.G. acknowledge the financial support of, "Tailor Graphene" ERC Advanced Grant (No. 321124) and the programme "Graphene physics in the time domain and applications to 3D optical memories", Aristeia II (No. 4470).